\documentclass[aps,prl,superscriptaddress,twocolumn,raggedbottom]{revtex4-2}
\usepackage{chngcntr}
\pdfoutput=1
\usepackage{dutchcal}
\DeclareMathAlphabet{\cmcal}{OMS}{cmsy}{m}{n}
\SetMathAlphabet{\cmcal}{bold}{OMS}{cmsy}{b}{n}
\newcommand{\resid}{\cmcal{D}}
\usepackage{graphicx}
\usepackage{xcolor}
\usepackage{siunitx}
\usepackage{amsmath,amssymb}
\usepackage[unicode=true,
  linktocpage,colorlinks=true,
  linkbordercolor={0.5 0.5 1},
  citebordercolor={0.5 1 0.5},
  linkcolor=blue]{hyperref}

\def\bea{\begin{equation}\begin{aligned}}
\def\eea{\end{aligned}\end{equation}}

\def\v{{\bf v}}
\def\k{{\bf k}}
\def\q{{\bf q}}

\newcommand{\callmgs}{\left\langle \callm \right\rangle}
\newcommand{\qmin}{q_\mathrm{min}}
\newcommand{\callm}{\mathcal{m}}
\newcommand{\dq}{d_q}
\newcommand{\gammamr}{\gamma_\mathrm{MR}}
\newcommand{\smcitations}{ThuillierEtAl2025, NilssonEtAl2025, NazaryanLevitov2024, LedwithEtAl2019, quartic-oscillator}

\makeatletter
\@addtoreset{equation}{superequation}
\makeatother

\begin{document}

\def\titlename{AC Fingerprints of 2D Electron Hydrodynamics: Superdiffusion and Drude Weight Suppression}
\title{\titlename}

\author{Davis Thuillier}
\affiliation{Department of Physics, University of California, Irvine, Irvine, CA 92697, USA}
\author{Thomas Scaffidi}
\email{tscaffid@uci.edu}
\affiliation{Department of Physics, University of California, Irvine, Irvine, CA 92697, USA}

\date{16 March 2026}

\begin{abstract}
Clean two-dimensional Fermi liquids are now known to exhibit an intermediate \emph{tomographic} regime, between ballistic and Navier--Stokes transport, caused by the anomalously slow relaxation of parity-odd multipolar deformations of the Fermi surface. Here we show that this anomaly extends to the dynamical realm. Starting from a microscopic numerical evaluation of the linearized electron--electron collision operator, we find that the finite-frequency nonlocal conductivity is controlled at low frequency by a single hydrodynamic pole, $\sigma(q,\omega)=\resid(q)/(i\omega+\eta_\star q^z)$, with dynamical exponent $z=4/3$ and superdiffusive viscosity $\eta_\star$. Remarkably, the pole residue itself is scale dependent and obeys $\resid(q)\sim q^{-\alpha}$ with $\alpha=1/3$, so the dynamical properties are described by two separate exponents rather than one. We interpret the residue suppression using a Krylov-chain description of current relaxation: as $q$ increases, the longest-lived quasinormal mode ceases to be a nearly pure current excitation and spreads over higher odd angular harmonics. Finally, we show that AC transport in narrow channels provides a direct experimental probe of these phenomena.
\end{abstract}

\maketitle

Hydrodynamics is a powerful framework for describing interacting many-body systems, and recent progress in ultra-clean metals has shown its relevance to electronic transport. In this regime, momentum-conserving electron collisions dominate over momentum-relaxing ones, allowing viscosity to control transport. This possibility has driven substantial theoretical \cite{Gurzhi1968, GurzhiEtAl1995, GURZHI1996497, TomadinEtAl2014, TorreEtAl2015, Svintsov2018, LevitovFalkovich2016, Scaffidi2017, Lucas2017, LucasFong2018, LucasHartnoll2018, MoessnerEtAl2018, HolderEtAl2019, KiselevSchmalian2019, KiselevSchmalian2019a, LevchenkoSchmalian2020, HuangLucas2021, LedwithEtAl2019, LedwithEtAl2019a, PhysRevB.99.235148, SternEtAl2022, HofmannDasSarma2022, HofmannGran2023, HongEtAl2020, KumarEtAl2022, diaz, NilssonEtAl2025, MoiseenkoEtAl2025, RostamiEtAl2025, AlvarezEtAl2025, KryhinEtAl2025, ThuillierEtAl2025, rwjf-lh2r, Musser, KHVESHCHENKO2024130006,PhysRevB.111.L161108,2026arXiv260306518M} and experimental \cite{deJongMolenkamp1995, KrishnaKumarEtAl2017, BandurinEtAl2016, BerdyuginEtAl2019, CrossnoEtAl2016, SulpizioEtAl2019, Aharon-SteinbergEtAl2022, PhysRevLett.129.087701, PalmEtAl2024, GeursEtAl2025, MollEtAl2016, NandiEtAl2018, BachmannEtAl2022, ZengEtAl2024} activity. It is by now well understood that viscous effects can be observed in certain metals through size-dependent transport, as captured by the non-local conductivity $\sigma(q)$.

A fundamental question raised by these developments is whether quantum effects can give rise to emergent hydrodynamic theories that differ qualitatively from the Navier--Stokes paradigm, as can occur in holographic theories~\cite{HartnollLucasSachdev2018,Blaise}. Low-dimensional systems also provide natural settings for such behavior. In quantum spin chains, for example, spin and energy transport can become superdiffusive \cite{Joel,LjubotinaEtAl2019, WeinerEtAl2020, PhysRevB.101.121106, TakeuchiEtAl2025,PhysRevX.15.010501}, with a dynamical exponent $z$ lying between the ballistic value $z=1$ and the diffusive value $z=2$.

Two-dimensional Fermi liquids provide another striking example. Because of the special form of Pauli blocking in two dimensions due to geometric constraints, parity-odd angular harmonics of the quasiparticle distribution near the Fermi surface relax parametrically more slowly than parity-even ones~\cite{GURZHI1996497,LedwithEtAl2019, LedwithEtAl2019a, KryhinEtAl2025, HofmannDasSarma2022, HofmannGran2023, HongEtAl2020, diaz, MoiseenkoEtAl2025, RostamiEtAl2025, AlvarezEtAl2025}. As a result, clean 2D Fermi liquids exhibit two distinct regimes of hydrodynamic behavior: at large length scales, 2D Fermi liquids behave like a conventional viscous fluid described by the Navier-Stokes equation with non-local transverse conductivity $\sigma(q)\sim q^{-2}$; at intermediate scales, however, they enter a so-called tomographic~\cite{LedwithEtAl2019} regime with fractional scaling $\sigma(q)\sim q^{-5/3}$ due to the long-lived odd modes~\cite{NazaryanLevitov2024}.

Here we are interested in how these hydrodynamic effects extend to the \emph{dynamical} realm, which is captured by the finite-frequency non-local conductivity $\sigma(q,\omega)$.
In the Navier-Stokes regime, going to finite frequency is fairly trivial: one finds a diffusive pole $\sigma(q,\omega) = 1 / (i\omega + \eta q^2)$, and the dynamical exponent $z=2$ could thus have been inferred directly from the static conductivity decaying as $\sigma(q,0) \sim q^{-2}$.
By contrast, we will show that the dynamical conductivity in the tomographic regime is considerably richer than its static limit suggests.
We find that the low-frequency transverse conductivity is governed by a pole of the form,
\bea\label{sigmaintro}
\sigma(q,\omega)\approx \frac{\resid(q)}{i\omega+\eta_\star q^z},
\eea
with dynamical exponent $z=4/3$ and superdiffusive viscosity $\eta_\star$. Crucially, the residue $\resid(q)$, which can be interpreted as a finite-$q$ Drude weight, is itself scale dependent:
\bea
\resid(q)\sim q^{-\alpha}, \qquad \alpha=\frac13.
\eea
The static scaling $\sigma(q,0)\sim q^{-5/3}$ therefore arises from the combination of a superdiffusive decay rate and an anomalously suppressed residue. The tomographic regime is thus characterized by two exponents, $z$ and $\alpha$, rather than one. 

We first compute the spectrum of the linearized electron--electron collision operator and verify the strong even--odd hierarchy of relaxation rates characteristic of 2D Fermi liquids. We then reformulate the finite-$q$ current dynamics as a diffusion-dissipation model on a Krylov chain, which yields the prediction of Eq.~\eqref{sigmaintro}. Next, we confirm these predictions with numerical calculations and we show how logarithmic corrections to the even decay rates govern the slow approach of the exponents $z$ and $\alpha$ to their asymptotic values. Finally, we show how to use AC transport in narrow channels to probe these effects experimentally.

While our focus is on superdiffusion and finite-$q$ Drude weight suppression, recent work has explored other dynamical consequences of the even-odd hierarchy, including cyclotron resonances \cite{MoiseenkoEtAl2025}, finite-frequency modifications of the Gurzhi dip \cite{rwjf-lh2r}, and collective modes in idealized tomographic spectra~\cite{HofmannDasSarma2022,KHVESHCHENKO2024130006,PhysRevB.111.L161108,2026arXiv260306518M}.

\emph{Boltzmann equation and collision spectrum.}---
We consider a clean two-dimensional Fermi liquid at $T\ll T_F$. 
A weak deformation of the Fermi--Dirac distribution is written as
$\delta f(\mathbf r,\mathbf k,t)=(-\partial n_F/\partial\varepsilon_\mathbf{k})\chi(\mathbf r,\mathbf k,t)$,
where $\chi$ denotes the nonequilibrium part of the distribution (with the Fermi-Dirac energy derivative factored out) and obeys the Boltzmann transport equation (BTE):
\begin{equation}
\partial_t \chi+\mathbf v\cdot\nabla_\mathbf r \chi+L[\chi]=e\,\mathbf v\cdot\mathbf E.
\label{eq:bte}
\end{equation}
Here $L$ is the linearized electron--electron collision operator, $\mathbf E$ is the applied electric field, and $\mathbf v=v_F(\cos\theta,\sin\theta)$ is the quasiparticle velocity on the Fermi surface (FS), with $\theta$ the polar angle along the FS~\footnote{We take $v_F=1$ throughout, except when giving numerical estimates.}

In order to solve this equation, the first step is to calculate the collision operator $L$.
We evaluate it numerically for a local repulsive interaction using the method introduced in \cite{ThuillierEtAl2025}; see the Supplemental Material (SM) \cite{supplement}\nocite{\smcitations}, Appendix A for details. 
Using rotational invariance, and in the limit $T\ll T_F$, the collision operator is fully characterized by decay rates $\gamma_m$ for the angular harmonics of the distribution $\chi(\theta)$ of quasiparticles at the Fermi surface
$
L[e^{im\theta}]=\gamma_m e^{im\theta},
$ and
with $\gamma_0=\gamma_1=0$ due to conservation of particle number and momentum.

\begin{figure}[t!]
    \centering
    \includegraphics[width=\linewidth]{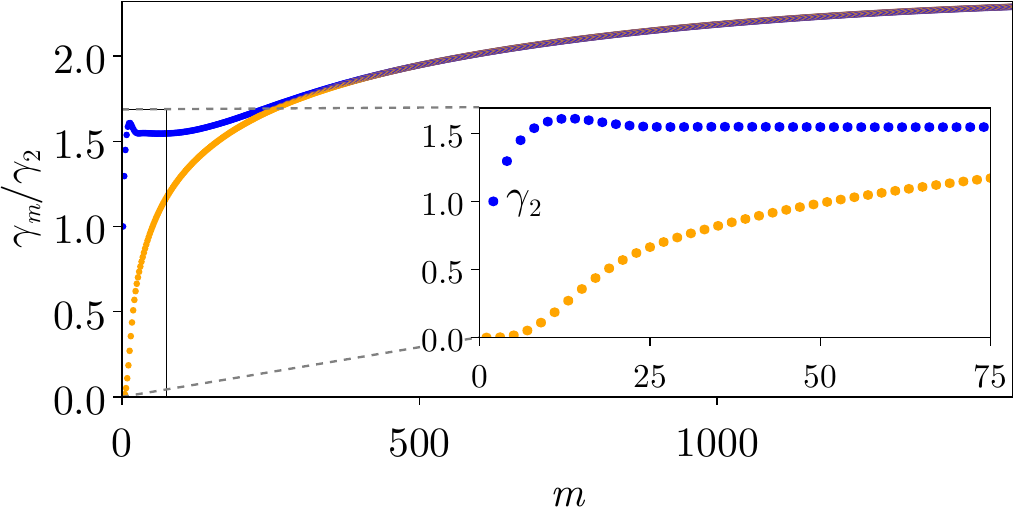}
    \caption{Spectrum of the linearized interparticle collision operator in the cylindrical harmonics basis, shown as the decay rates $\gamma_{m,\text{even}}$ in blue and $\gamma_{m,\text{odd}}$ in orange at $T/T_F = 0.005$. One can observe the different behavior of the even and odd rates for $m \leq m_\text{max} \approx 12$ as discussed around Eqs.~\ref{eq:hofmann-even} and \ref{eq:hofmann-odd}.
    }
    \label{fig:spectrum}
\end{figure}

The resulting spectrum, shown in \autoref{fig:spectrum}, exhibits the expected $\gamma_{m,\text{even}} \gg \gamma_{m,\text{odd}}$ hierarchy already mentioned in the introduction~\cite{LedwithEtAl2019, LedwithEtAl2019a, KryhinEtAl2025}. Rates for even harmonics are approximately constant in $m$: they increase only slowly with $m$ up to $m_{\max}\sim (T/T_F)^{-1/2}$ after which they saturate, see inset of Fig.~\ref{fig:spectrum}.
Our results are in good agreement with the analytical prediction of Ref.~\cite{NilssonEtAl2025}:
\bea
\gamma_{m,\mathrm{even}} =
\gamma_2 F[m],
\label{eq:hofmann-even}
\eea
with $\gamma_2 \propto T^2/T_F$ and where
$
F[m]
$ is a slowly increasing function of $m$ such that $F[2]=1$ and $F[m\to\infty]\sim \frac12\log m$ (see SM App. B and C for details).

Decay rates of odd harmonics grow as $m^4$ up to $m_{\max}$, after which they saturate. For $m\ge 3$, Ref.~\cite{NilssonEtAl2025} obtained
\bea
\gamma_{m,\mathrm{odd}}
=
\gamma_3\left(\frac{m}{3}\right)^4
\label{eq:hofmann-odd}
\eea
with $\gamma_3 \propto T^4/T_F^3$ (see SM App. B for full expression).
When performing calculations, we will use our numerically obtained spectra for $T/T_F \geq 0.005$, and the analytic expressions \eqref{eq:hofmann-even} and \eqref{eq:hofmann-odd} when studying lower $T/T_F$.

\emph{Diffusion-dissipation dynamics on the Krylov chain.}---
To understand the pole structure of the transverse $\sigma(q,\omega)$, it is useful to start with the real-time relaxation of a current flowing along $x$ and modulated along $y$,
$
\mathbf{j}(y,t)=e^{iqy}j_q(t)\,\hat{\mathbf{x}},
$
with initial condition $j_q(0)=1$.

The initial current corresponds in the BTE to the initial condition $\chi(y,\theta,t=0)=e^{i q y} \cos\theta$. Omitting the $e^{i q y}$ factor henceforth, we expand $
\chi(\theta,t)=\sum_{m\ge 1}\varphi_m(t)\,f_m(\theta),
$
with
\begin{equation}
f_m(\theta)=
\begin{cases}
\cos(m\theta), & m \text{ odd},\\
-i\sin(m\theta), & m \text{ even},
\end{cases}
\end{equation}
where the coefficients $\varphi_m(t)$ may be viewed as a wavefunction on a semi-infinite one-dimensional ``Krylov'' chain~\cite{PhysRevX.9.041017} whose site index $m$ labels angular harmonics of the FS deformation (see SM App. D). 

The BTE then takes the form
\bea
\partial_t \varphi_m
=
\frac{q}{2}\left(\varphi_{m-1}-\varphi_{m+1}\right)
-\gamma_m\varphi_m,
\qquad m\ge 1,
\label{eq:krylov-chain}
\eea
with $\varphi_0\equiv 0$. The $q$-dependent term generates hopping along the chain, while the collision operator produces site-dependent decay. We focus on $m$ below $m_\text{max}$, in which case $\gamma_{2n}\gg\gamma_{2n-1}$, and the even sites may thus be eliminated adiabatically: $
\varphi_{2n}\simeq \frac{q}{2\gamma_{2n}}
\left(\varphi_{2n-1}-\varphi_{2n+1}\right).
$
Approximating for now the even rates as constant ($\gamma_{2n}\approx \gamma_2$), the dynamics of the odd modes reads:
\bea
\partial_t \varphi_1 &= \frac{q^2}{4\gamma_2} (\varphi_3 - \varphi_1) \\
\partial_t \varphi_{m \geq 3}
&=
\frac{q^2}{4\gamma_2}
\left(
\varphi_{m-2}-2\varphi_m+\varphi_{m+2}
\right)
-\gamma_m\varphi_m, 
\label{eq:odd-discrete-diffusion-simple}
\eea
with initial condition $\varphi_m(0) = \delta_{m,1}$ and where the current is given by $j_q(t) = \varphi_1(t)$.

For $q \ll \qmin$ (with $\qmin \equiv \sqrt{\gamma_2 \gamma_3}$), the second line of Eq.~\ref{eq:odd-discrete-diffusion-simple} is solved at leading order by $\varphi_{m\geq3} \approx 0 $, in which case the current is obtained from the first line as:
\bea
j_q(t)\sim e^{-\eta q^2 t}
\qquad\text{(Navier--Stokes)},
\label{eq:jqNS}
\eea
with $\eta = 1 / (4 \gamma_2)$ the viscosity. 
This is the familiar Navier-Stokes regime.

By contrast, in the tomographic regime for $q \gg \qmin$, one needs to solve the full set of equations in \eqref{eq:odd-discrete-diffusion-simple}, which is a discrete diffusion-dissipation model with diffusion constant $\dq\equiv q^2/\gamma_2$ and dissipation rates set by $\gamma_{m,\text{odd}}$.
Since the wavefunction now spreads over many odd sites, we take the continuum limit $m \to \callm \in \mathbb{R}$.
Eq.~\ref{eq:odd-discrete-diffusion-simple} then becomes $-\partial_t \varphi(\callm,t)=H\varphi(\callm,t) $ with
\bea
H=-\dq\,\partial_{\callm}^2+V(\callm),
\label{eq:Schr}
\eea
a 1D Schr\"odinger operator: the diffusion term $-\dq\,\partial_{\callm}^2$ plays the role of a kinetic energy, while dissipation acts as a confining potential, $V(\callm)=\gamma_{m,\mathrm{odd}}\propto \gamma_3 |\callm|^p$ with $p=4$.

After diagonalizing the Hamiltonian,
\bea
H\phi_\mu(\callm)=\epsilon_\mu\,\phi_\mu(\callm),
\eea
with eigenvalues ordered as $0<\epsilon_0<\epsilon_1<\cdots$, one has
\bea\label{jqfromspec}
j_q(t)=\sum_\mu |\phi_\mu(\callm=0)|^2 e^{-\epsilon_\mu t}
\eea
since $j_q(t) = \varphi(\callm=0,t)$ in the continuum limit becomes the return amplitude of a wavepacket initially localized at the origin: $\varphi(\callm,t=0) = \delta(\callm)$.
\footnote{Upon taking the continuum limit, the first Krylov site---corresponding to the current--- maps to $\callm=0$, and the original semi-infinite chain may be unfolded to the full real line provided one restricts to wavefunctions even under $\callm\to-\callm$.}.

For $p=4$, $V(\callm) \propto m^4$, and $H$ describes a quantum-mechanical quartic oscillator with low-energy eigenstates $\phi_\mu(\callm)$ forming a discrete set of localized states with a characteristic extent $\callmgs$ set by balancing kinetic and potential energy:
\bea \label{conditionform}
\dq/\callmgs^2\sim \gamma_3\callmgs^p \implies \callmgs\sim\left(\dq/\gamma_3\right)^{1/(p+2)}.
\eea
Similarly, the corresponding eigenvalues scale as $\epsilon_\mu \sim \gamma_3 \left(\dq/\gamma_3\right)^{p/(p+2)}$.

At early times, the initial wave packet is localized near the origin, the confining potential is negligible, and the wave packet simply diffuses, so its extent grows as $\callm(t)\sim \sqrt{\dq t}$. This gives a power-law decay of the current
\bea
j_q(t)\sim \frac{1}{\sqrt{\dq t}}, \qquad t \ll t(q)
\label{eq:power-law-decay}
\eea
which continues until a time $t(q) \sim \callmgs^2/\dq$, when the spreading of the wavepacket saturates when $\callm(t) \sim \callmgs$.

At later times, the fact that $H$ is gapped~\footnote{As discussed in the SM, Appendix E, for a quartic oscillator, $\epsilon_1 / \epsilon_0 \approx 7$.} justifies keeping only the ``ground state'' $\phi_0(\callm)$ in \eqref{jqfromspec}:
\bea
j_q(t)\simeq \resid(q)e^{-\gamma(q)t}, \qquad t \gg t(q)
\label{eq:jq}
\eea
with $\gamma(q)\equiv \epsilon_0(q)$ and $\resid(q)\equiv |\phi_0(\callm=0)|^2$. Using the scaling found above for the spectrum of $H$, we obtain
\bea
\gamma(q)
\sim
\gamma_3\left(\frac{q}{\qmin}\right)^z,
\qquad
z=\frac{2p}{p+2}=\frac{4}{3}.
\label{eq:gammaq}
\eea

Likewise, since the ground state is spread over $\sim \callmgs$ sites, one finds $\resid(q) \sim \callmgs^{-1}$ and thus
\bea
\resid(q)
\sim
\left(\frac{q}{\qmin}\right)^{-\alpha},
\qquad
\alpha=\frac{2}{p+2}=\frac{1}{3}.
\label{eq:residq}
\eea

Overall, the real-time dynamics in the tomographic regime differs qualitatively from Navier--Stokes (Eq.~\ref{eq:jqNS}) because the relevant eigenmode $\phi_0(\callm)$ is spread over a parametrically large number of higher odd harmonics, rather than being an almost purely dipolar excitation. Consequently, $j_q(t)$ first decays as $1/\sqrt{t}$ while the wavepacket diffuses across the full extent of this mode, and only then crosses over to exponential decay, $j_q(t)\sim \resid(q)e^{-\gamma(q)t}$, with a parametrically small prefactor $\resid(q)\ll 1$. 
In frequency space, this will becomes the suppressed pole residue of the conductivity.
(Before moving to frequency space, we note that the two exponents $z$ and $\alpha$ also dictate the real-space current-current correlator, with the predicted scaling form
$
C(x,t)=t^{-\frac{1-\alpha}{z}}\,f\!\left(x/t^{1/z}\right)$
with $\frac{1-\alpha}{z}=\frac12.
$)

\begin{figure}[t!]
    \centering
    \includegraphics[width=\linewidth]{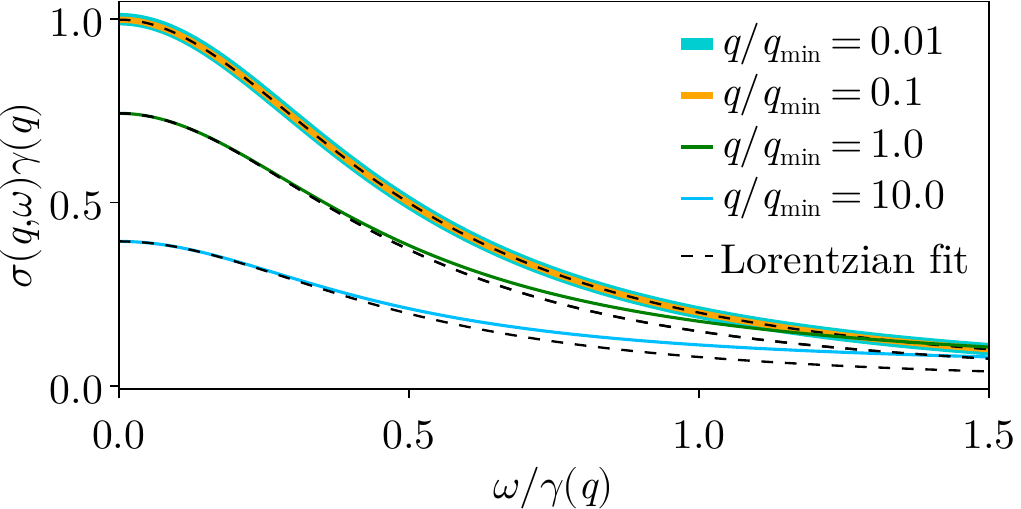}
    \caption{Real part of the conductivity $\sigma(q,\omega)$ at $T/T_F=0.005$. The low-frequency line shape is well fit by a Lorentzian, $\sigma(q,\omega)=\resid(q)/(i\omega+\gamma(q))$, from which we extract the decay rate $\gamma(q)$. Rescaling the vertical axis by $\gamma(q)$ and the horizontal axis by $1/\gamma(q)$ would collapse the curves if the residue were constant. This collapse is observed for the two curves with $q\ll \qmin$ (which are in the Navier-Stokes regime), while the two curves with $q\gtrsim \qmin$ (which are in the tomographic regime) are progressively suppressed, indicating a decreasing residue. Note also that the latter two curves exhibit a slower high-frequency decay than predicted by the low-frequency Lorentzian fit, due to contributions from additional poles with higher decay rates.}
    \label{fig:frequency-dependent-conductivity}
\end{figure}

\begin{figure*}[t!]
    \centering
    \includegraphics[width=\linewidth]{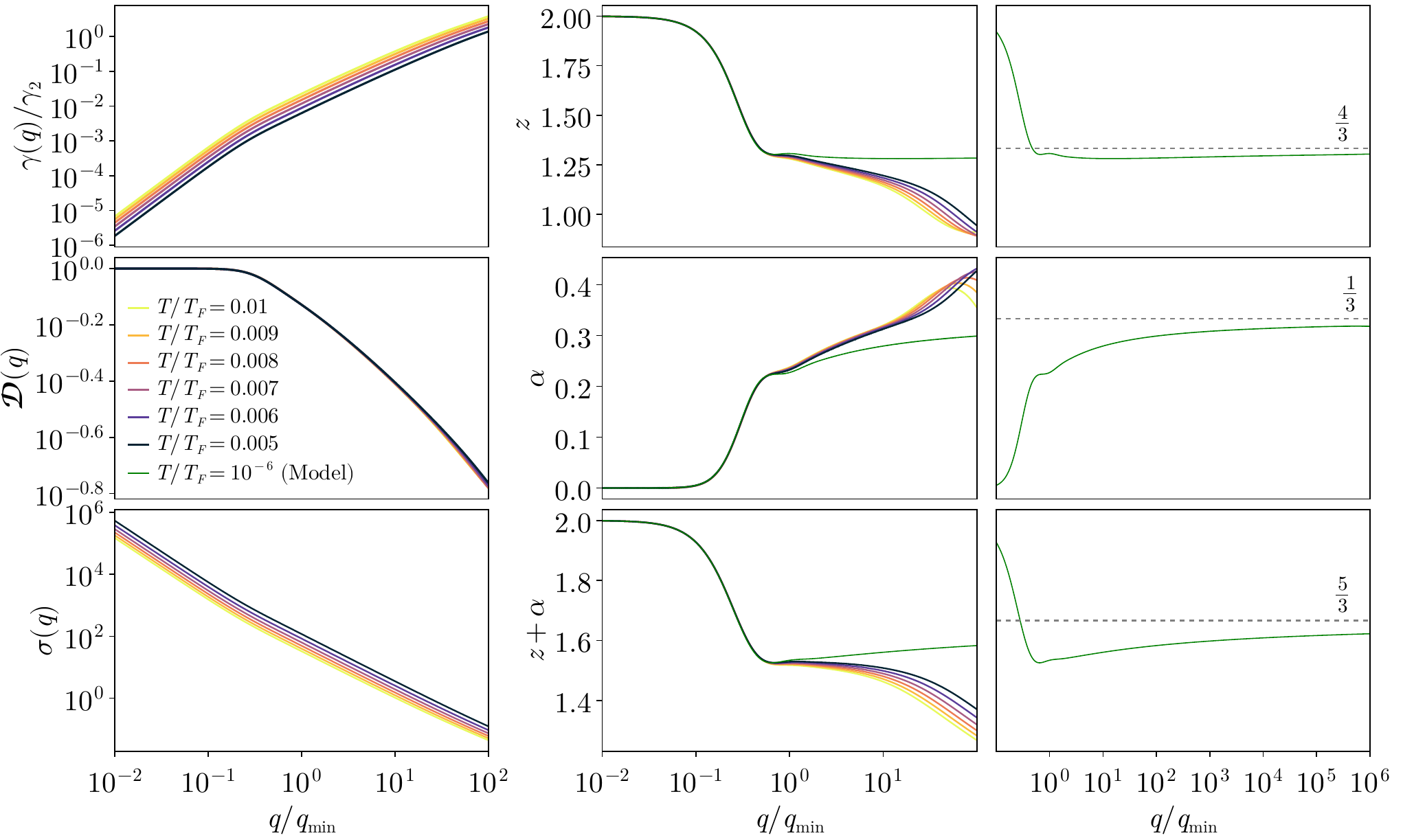}
    \caption{(Left column) Extracted decay rate $\gamma(q)$, pole residue $\resid(q)$, and DC conductivity $\sigma(q,\omega=0)$ for a range of temperatures.  (Middle column) Local exponents calculated by logarithmic derivatives, i.e., $z(q)\equiv d\ln\gamma/d\ln q$ and $\alpha \equiv d\ln\resid/d\ln q$, plotted versus $q/\qmin$. (Right column) Local exponents shown over a larger range of $q/\qmin$, for $T/T_F = 10^{-6}$, demonstrating the slow approach to their asymptotic values.}
    \label{fig:conductivity}
\end{figure*}

\emph{AC conductivity.}---
Up to an overall prefactor of the ``bulk'' Drude weight $D_0$ which we will omit henceforth ($D_0 \equiv n e^2/m^*$ with $n$ the carrier density and $m^*$ the effective mass)~\footnote{Note that systems without Galilean invariance, like graphene with its linear dispersion, can already have a modified value of $D_0$ due to interaction effects~\cite{DrudeNongalilean}. The finite-$q$ Drude weight suppression we discuss below, however, is strictly different from this, and would also occur for Galilean-invariant systems.}, the transverse conductivity is
\bea
\sigma(q,\omega)=\int_0^\infty dt\, j_q(t)e^{-i\omega t}.
\label{eq:sigma-from-j}
\eea
In the Navier--Stokes regime, plugging \eqref{eq:jqNS} in this integral gives the familiar diffusive pole,
\bea
\sigma(q,\omega)=\frac{1}{i\omega+\eta q^2}
\qquad\text{(Navier--Stokes)}.
\label{eq:diffusive-scaling}
\eea
In the tomographic regime, inserting \eqref{eq:jq} instead yields
\bea
\sigma(q,\omega)
\approx
\frac{\resid(q)}{i\omega+\gamma(q)}
=
\frac{\tilde q^{-\alpha}}{i\omega+\eta_\star q^z},
\label{eq:single-pole}
\eea
with $\tilde q\equiv q/\qmin$ and where
$
\eta_\star \equiv \gamma_3/\qmin^z
$
is the superdiffusive version of the viscosity.
We thus find that the hydrodynamic pole is described by two exponents, the dynamical exponent $z$ describing the superdiffusive scaling of the decay rate $\gamma(q)$, and the exponent $\alpha$ controlling the suppression of the pole residue $\resid(q)$. This pole residue can be interpreted as a finite-$q$ Drude weight $\resid(q)$, given in units of $D_0$. In accordance with the $f$-sum rule $\int_0^\infty d\omega \mathrm{Re}[\sigma(q,\omega)] = \frac{\pi}2$ (see SM, Appendix E), the suppression of the low frequency peak is compensated by a redistribution of the spectral weight to higher frequencies, as can be seen in Fig.~\ref{fig:frequency-dependent-conductivity}.

Another formulation of our main result Eq.~\ref{eq:single-pole} is the fractional Stokes equation
\bea
\partial_t \mathbf{j}
=
-\eta_\star |\nabla|^z \mathbf{j}
+
\qmin^\alpha |\nabla|^{-\alpha}\mathbf{E},
\label{eq:stokes}
\eea
where the gradient is taken transverse to $\mathbf{E}$. This equation applies over the tomographic range $ \qmin \ll q \ll  \gamma_2/v_F$.

\emph{Numerical confirmation.}---
We now compare these predictions with $\sigma(q,\omega)$ computed numerically from the full spectrum of decay rates using the continued-fraction representation of Ref.~\cite{NazaryanLevitov2024}. Figure~\ref{fig:frequency-dependent-conductivity} shows a representative set of $\mathrm{Re}[\sigma]$ vs $\omega$ line shapes. For each fixed $q$, we perform a low-frequency fit to the form $
\sigma(q,\omega)=\frac{\resid(q)}{i\omega+\gamma(q)}
$ and extract $\gamma(q)$ and $\resid(q)$.
The extracted quantities are shown in Fig.~\ref{fig:conductivity}. As expected, one recovers the Navier--Stokes regime for $q\ll \qmin$, with $\gamma(q) \sim q^2$ and $\resid(q) =1$. For $q\gtrsim \qmin$, we observe a crossover to the tomographic regime: the decay rate crosses over to $\gamma(q)\sim q^z$ with $z$ approaching $4/3$, and the residue starts to decay as $\resid(q)\sim q^{-\alpha}$ with $\alpha$ approaching $1/3$.

We observe that the approach with increasing $q/\qmin$ of the ``local exponents'' $z(q)$ and $\alpha(q)$ towards their asymptotic value is slow, see Fig.~\ref{fig:conductivity} right column.
 We attribute this to the slow increase with $m$ of the even rates (see Eq.~\ref{eq:hofmann-even}), which we had neglected and now reintroduce. Since $\gamma_{m,\text{even}} \propto F[m]$ grows as $\frac12\log m$ for $1 \ll m \ll m_\text{max}$, the effective diffusion constant along the Krylov chain becomes $\callm$ dependent, $\dq\sim q^2/(\gamma_2 F[\callm])$. Equation~\eqref{conditionform} is then replaced by
$
F[\callm]\callm^{p+2}\sim (q/\qmin)^2.
$
This motivates an effective exponent
\bea
p_\mathrm{eff}[\callm]
&\equiv
\frac{d\log\!\big(F[\callm]\callm^{p}\big)}{d\log \callm}
\sim
p+\frac{1}{\log (2\callm)+\gamma},
\label{eq:peff}
\eea
with $\gamma$ the Euler--Mascheroni constant (see SM, Appendix C). Equivalently, the observed exponents may be viewed as scale dependent, with $\alpha(q)=2/[p_\mathrm{eff}(q)+2]$ and $z(q)=p\alpha(q)$, where $p_\mathrm{eff}(q)\equiv p_\mathrm{eff}[\callm(q)]$ and $\callm(q)$ is determined by Eq.~\ref{conditionform}. As $q/\qmin$ increases, $\callm(q)$ increases as well, reducing the logarithmic correction and driving $p_\mathrm{eff}\to4$, $\alpha\to1/3$, and $z\to4/3$. In practice, however, the accessible range of $q/\qmin$ is upper bounded by the crossover to the ballistic regime, so experimentally measured exponents will typically remain below their asymptotic values, an important point when comparing with tomographic-regime experiments such as Ref.~\cite{ZengEtAl2024}.

\emph{AC conductance of a narrow channel.}---
A natural way to probe $\sigma(q,\omega)$ experimentally is through AC transport in a channel of width $W$. In the tomographic regime, where $W\ll \qmin^{-1}$ so that the relevant transverse wave numbers satisfy $q\sim 1/W\gg \qmin$, the low-frequency channel conductance inherits the same pole structure:
\bea
\frac{G(W,\omega)}{W}
\equiv
\frac{\resid(W)}{i\omega+\gamma(W)}
\sim
\frac{\qmin^\alpha W^\alpha}{i\omega+\eta_\star W^{-z}}.
\label{eq:channel-pole}
\eea
This ``finite-width Drude peak'' provides a direct way to measure $z$ and $\alpha$ separately by considering channels of varying width. The frequency width of the peak scales as $W^{-z}$, while its height scales as $G(\omega=0)/W \sim W^{z+\alpha}$. Equivalently, the total spectral weight under the peak scales as $W^\alpha$, giving a direct probe of the residue suppression.

\begin{figure}[t!]
    \centering
    \includegraphics[width=\linewidth]{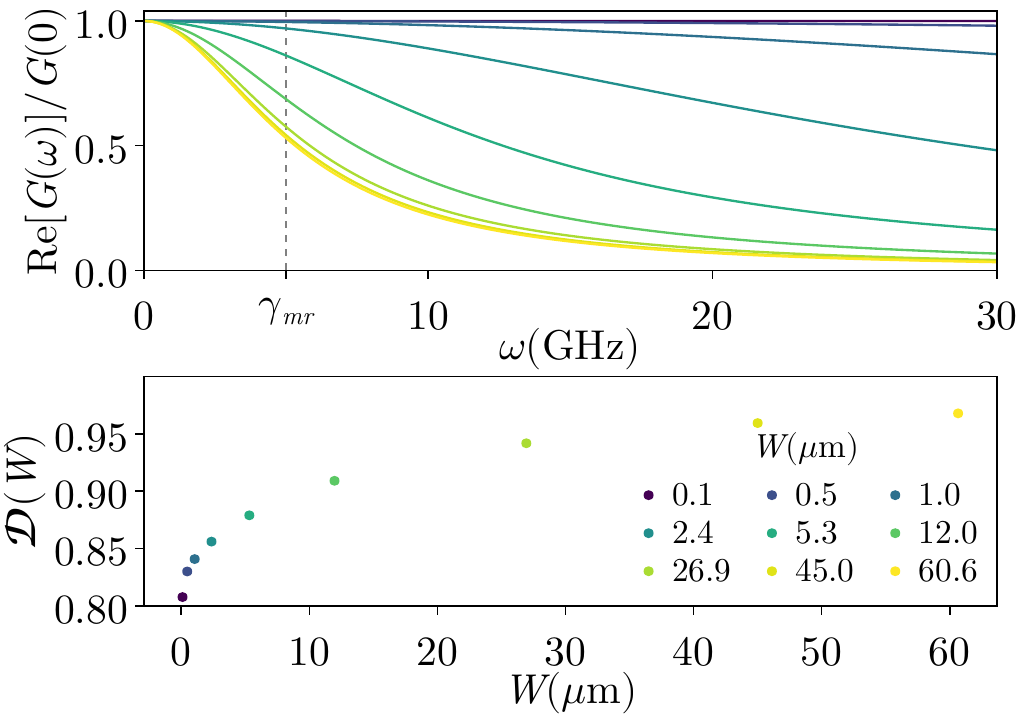}
    \caption{Top: Channel conductance for several widths $W$, computed at $T/T_F=0.005$ with $v_F=10^6\,\mathrm{m/s}$ and with an overall scale of the collision operator chosen so that $\gamma_3=5\,\mathrm{GHz}$. Momentum relaxation with $\gammamr=5\,\mathrm{GHz}$ is also included. Bottom: Pole residue extracted by fitting the low-frequency channel conductance to $G(\omega)/W=\resid/(i\omega+\gamma)$, showing residue suppression for narrow channels.}
    \label{fig:graphene-channel-conductance}
\end{figure}

We verify this behavior numerically in Fig.~\ref{fig:graphene-channel-conductance} (see the SM, Appendix F for details of the channel calculation), where the bottom panel shows the suppression of the Drude weight as the channel becomes narrower.
To model a more realistic experimental setting, we included a finite momentum-relaxation rate $\gammamr$ due to impurities or phonons via $\gamma_{m\ge 1}\to \gamma_{m\ge 1}+\gammamr$.
The scale $\gammamr$ sets a lower bound on the channel decay rate $\gamma(W)$ in Eq.~\ref{eq:channel-pole}; see Fig.~\ref{fig:graphene-channel-conductance}, top panel. Equivalently, the associated momentum-relaxing mean free path contributes to setting an upper width cutoff above which hydrodynamic effects become negligible.
In Fig.~\ref{fig:graphene-channel-conductance} we choose an aspirationally low value, $\gammamr=5\,\mathrm{GHz}$, to illustrate a best-case scenario: the effect could then be resolved from Drude peaks with widths of order tens of GHz in devices tens of microns wide.
Notably, this is not too far from the setup of Ref.~\cite{Lk}, in which the AC response of high-quality graphene was extracted, with a decay rate we infer to be around $ 30\,\mathrm{GHz}$.
For larger momentum-relaxation rates, e.g. $\gammamr\sim$ a few hundred $\mathrm{GHz}$, the relevant measurements would be pushed into the lower-THz range and would require devices only a few microns wide. Encouragingly, the conductivity of graphene samples can be measured in that frequency range~\cite{doi:10.1126/science.aat8687}.

One may also work at fixed width and vary temperature within the tomographic range. Since $\qmin\propto T^3$ and $\eta_\star\propto T^{4-3z}$, the peak height is predicted to scale as
$
G(\omega=0)/W\sim T^{3(z+\alpha)-4} 
$
while the frequency width of the peak scales as
$
\gamma\sim T^{4-3z} 
$. 
At the asymptotic values $z=4/3$ and $\alpha=1/3$, the peak height is linear in $T$ (as predicted in Ref.~\cite{KryhinEtAl2025}) while the peak width is $T$ independent, although the logarithmic drift of exponents discussed above implies noticeable deviations from these powers. The total Drude spectral weight increases with temperature as $T^{3\alpha}$, which is $T$-linear for $\alpha \to 1/3$. Overall, the $T$-linear increase of the DC conductance in the tomographic regime~\cite{KryhinEtAl2025} thus results from a $T$-linear Drude weight rather than from a temperature-dependent decay rate.

As discussed in the End Matter, at higher frequencies a different regime appears: the in-phase current becomes confined to boundary layers near the channel walls~\cite{MoessnerEtAl2018}, whereas the bulk becomes purely inductive. For a dynamical exponent $z$, the boundary-layer width is expected to scale as
$
\delta\sim \left(\frac{\eta}{\omega}\right)^{1/z}.
$
We confirmed this prediction numerically, showing a crossover of the form:
\bea\label{eq:crossoverdelta}
\delta \sim
\begin{cases}
(\eta/\omega)^{1/2}, & \text{Navier--Stokes } (\omega \ll \gamma_3),\\[3pt]
(\eta_\star/\omega)^{3/4} & \text{Tomographic } (\omega \gg \gamma_3).
\end{cases}
\eea

\emph{Conclusion}.---
We have shown that the finite-frequency nonlocal conductivity of clean 2D Fermi liquids in the tomographic regime is controlled by a single hydrodynamic pole,
$
\sigma(q,\omega)\sim \frac{\resid(q)}{i\omega+\eta_\star q^z},
$
with $z=4/3$ and a scale-dependent residue $\resid(q)\propto q^{-\alpha}$ with $\alpha=1/3$. 
The suppressed residue is an intrinsic part of the hydrodynamic structure, reflecting the delocalization of the slowest quasinormal mode over many odd angular harmonics. 
The effects reported here are not restricted to genuinely two-dimensional materials such as graphene, but are also expected in bulk metals with quasi-two-dimensional cylindrical Fermi surfaces, for which the skin effect may offer a direct experimental probe~\cite{PhysRevX.14.011018}.
Finally, we note that the dynamical exponent $z=4/3$ has also appeared in a phenomenological scaling theory of the strange-metal phase~\cite{PhysRevB.91.155126}.

\begin{acknowledgments}
\emph{Acknowledgments}.---
This work was supported by the U.S. Department of Energy, Office of Science, Office of Basic Energy Sciences under Early Career Research Program Award No.~DE-SC0025568. DT gratefully acknowledges support from the Eddleman Quantum Institute graduate fellowship. Simulation codes for generating the collision-operator spectrum are available on Github as Ludwig.jl v0.2.1 \cite{Ludwig}. We gratefully acknowledge Graham Baker, Jack Farrell, Blaise Gout\'eraux, Sean Hartnoll, Brad Ramshaw, Javier Sanchez-Yamagishi, and Christopher Yang for valuable comments on the manuscript.
\end{acknowledgments}

\bibliography{main}

\clearpage
\onecolumngrid
\begin{center}
\vspace{0.2cm}
{\large\bf End Matter}
\end{center}
\twocolumngrid

\begin{figure}[t!]
    \centering
    \includegraphics[width=\linewidth]{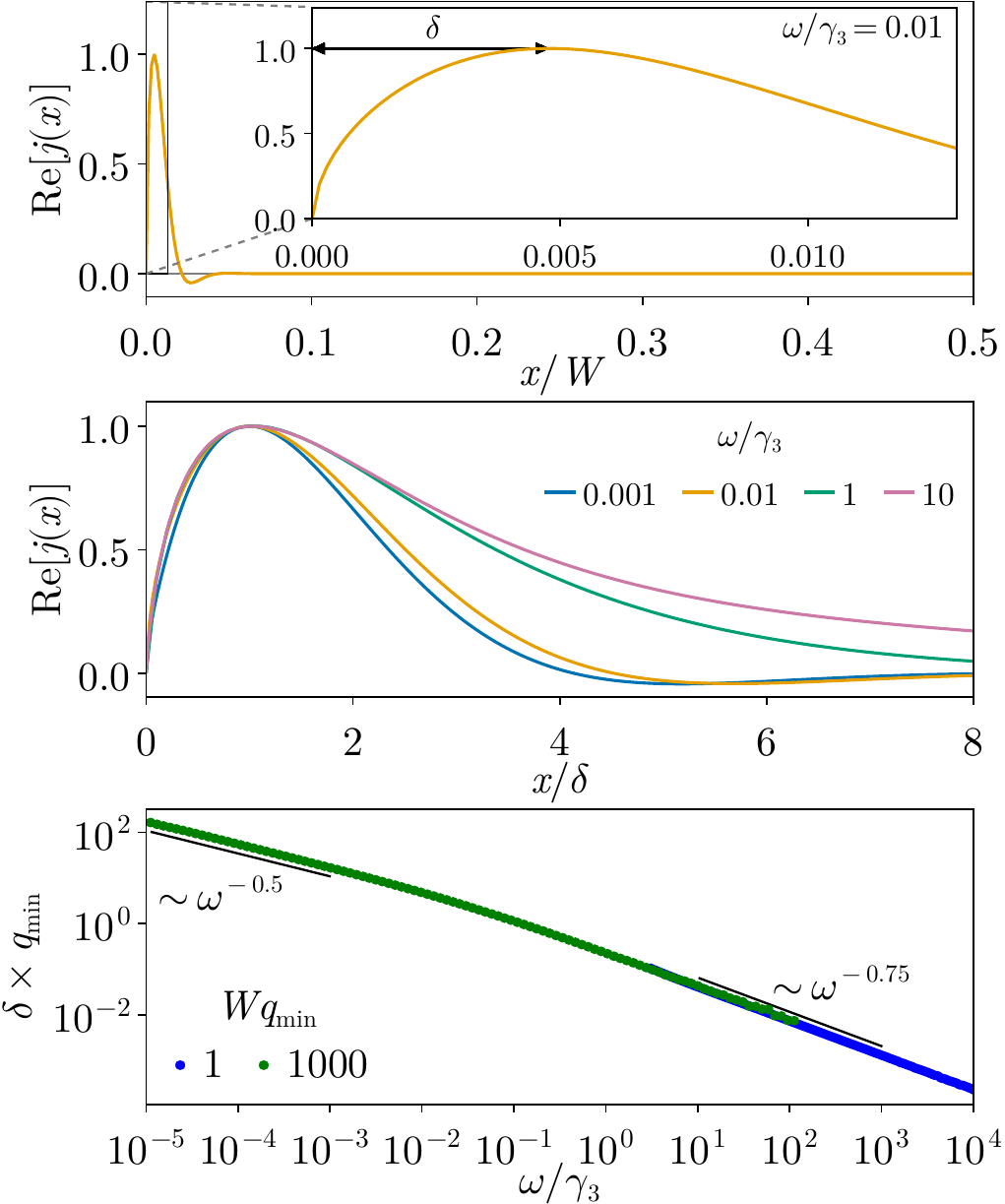}
    \caption{Top: Representative current profile at $T/T_F=10^{-5}$ in a channel of width $W=1000\,q_\mathrm{min}^{-1}$. The maximum defines the boundary-layer width $\delta$. Middle: The current shows the usual exponential damping with $x/\delta$ in the NS regime ($\omega \ll \gamma_3$), but seems to penetrate further into the bulk in the tomographic regime ($\omega \gg \gamma_3$), presumably due to the non-analytic-in-$q$ form of $\sigma$. Bottom: frequency dependence of $\delta$, showing the crossover from $\delta\sim \omega^{-1/2}$ for $\omega\ll\gamma_3$ to $\delta\sim \omega^{-3/4}$ for $\omega\gg\gamma_3$.}
    \label{fig:skin-depth}
\end{figure}

\emph{Current profiles and boundary layers.}---
The current profile in a channel $j(x)$ in a channel $x \in [0,W]$ can be obtained from $\sigma(q,\omega)$ following Ref.~\cite{LedwithEtAl2019}; the derivation is summarized in the SM, Appendix F. In the low-frequency tomographic regime, the profile takes the form $j(x)\sim x^{1/3}(W-x)^{1/3}$ already discussed in Ref.~\cite{LedwithEtAl2019}. At higher frequency, when $\omega\gg \eta_\star W^{-z}$, the real part of the current becomes confined to boundary layers of width $\delta$ near the walls (see Fig.~\ref{fig:skin-depth}, top and middle), while the bulk response is predominantly inductive~\cite{Lk}. 
Fig.~\ref{fig:skin-depth} bottom confirms the crossover predicted by Eq.~\ref{eq:crossoverdelta} in the frequency scaling of the boundary-layer width, from the Navier--Stokes regime ($\delta \gg \qmin^{-1}$, or $\omega \ll \gamma_3$) to the tomographic regime ($\delta \ll \qmin^{-1}$, or $\omega \gg \gamma_3$).
Appendix F of the SM further shows how the total channel conductance depends on frequency and temperature in these two regimes.

\renewcommand{\sectionautorefname}{Appendix}
\setlength{\belowcaptionskip}{0pt}
\onecolumngrid

\begin{center}
    \large\textbf{Supplemental material to ``\textit{\titlename}''}
\end{center}

\appendix
\counterwithout{equation}{section}
\counterwithout{figure}{section}
\counterwithout{table}{section}

\setcounter{equation}{0}
\setcounter{figure}{0}
\setcounter{table}{0}

\renewcommand{\theequation}{S\arabic{equation}}
\renewcommand{\thefigure}{S\arabic{figure}}
\renewcommand{\thetable}{S\arabic{table}}

\renewcommand{\theHequation}{S\arabic{equation}}
\renewcommand{\theHfigure}{S\arabic{figure}}
\renewcommand{\theHtable}{S\arabic{table}}

\setcounter{secnumdepth}{3}


\section{Computing the spectrum of the collision operator $L$}

Applying the method we introduced in \cite{ThuillierEtAl2025}, we performed a direct calculation of the linearized electron-electron collision operator for a single parabolic band with short-range repulsion. 
The collision integral kernel $\cmcal{L}(\k_1, \k_2)$ can be expressed in terms of energy-angle coordinates. 
Due to rotational symmetry and mirror symmetry about $(\k + \k^\prime) / 2$, the kernel is an even function of the angular difference, $\theta = \theta_1 - \theta_2$ between the momenta.
For each harmonic, we can then integrate out this angular dependence to define $\cmcal{L}_m(\varepsilon, \varepsilon^\prime) = \int d\theta \cos(m \theta) \cmcal{L}(\varepsilon, \varepsilon^\prime, \theta)$. In practice, momenta are sampled on an energy-angle grid with $n_\varepsilon$ points sampled along the energy direction and $n_\theta$ points sampled in angle.
$\cmcal{L}_m$ is thus represented by an $n_\varepsilon \times n_\varepsilon$ matrix. 

We define the rate $\gamma_m$ to be the minimal magnitude eigenvalue of the $\cmcal{L}_m$ matrix.
It is possible to replace the scalar rates in the continued fraction method of \citet{NazaryanLevitov2024} with the $\cmcal{L}_m$ matrices and evaluate the nonlocal conductivity by taking the inner product $\langle v(\varepsilon)|\hat{\Gamma}_q|v(\varepsilon) \rangle$ where $\hat{\Gamma}_q$ denotes the corresponding \emph{operator continued fraction}.
However, larger eigenvalues in each block correspond to short-lived eigenfunctions having weak overlap with the rigid chemical potential shift we consider, and thus do not contribute substantially to the conductivity.

\emph{Modeling the spectrum at low-$T$}.---
As temperature is lowered, the error of all rates scales as $1 / n_\varepsilon$ for fixed angular resolution (standard Riemannian integration error). 
However, the error of the lowest odd rates ($m = 3, 5, 7, \dots$) increases relative to the rest of the spectrum as temperature is lowered, demanding $n_\varepsilon$ be increased to achieve the same level of convergence.
The computation time for $L$ is $\cmcal{O}((n_\theta n_\varepsilon)^3)$, so the convergence of the odd rates ultimately limits the minimal temperature for which $L$ can be computed in fixed time.  
In order to compute dynamics at lower temperatures than achievable through direct computation of $L$, we turn to modeling the spectrum at small $m$.

\begin{figure}[b!]
    \centering
    \includegraphics[width=0.5\linewidth]{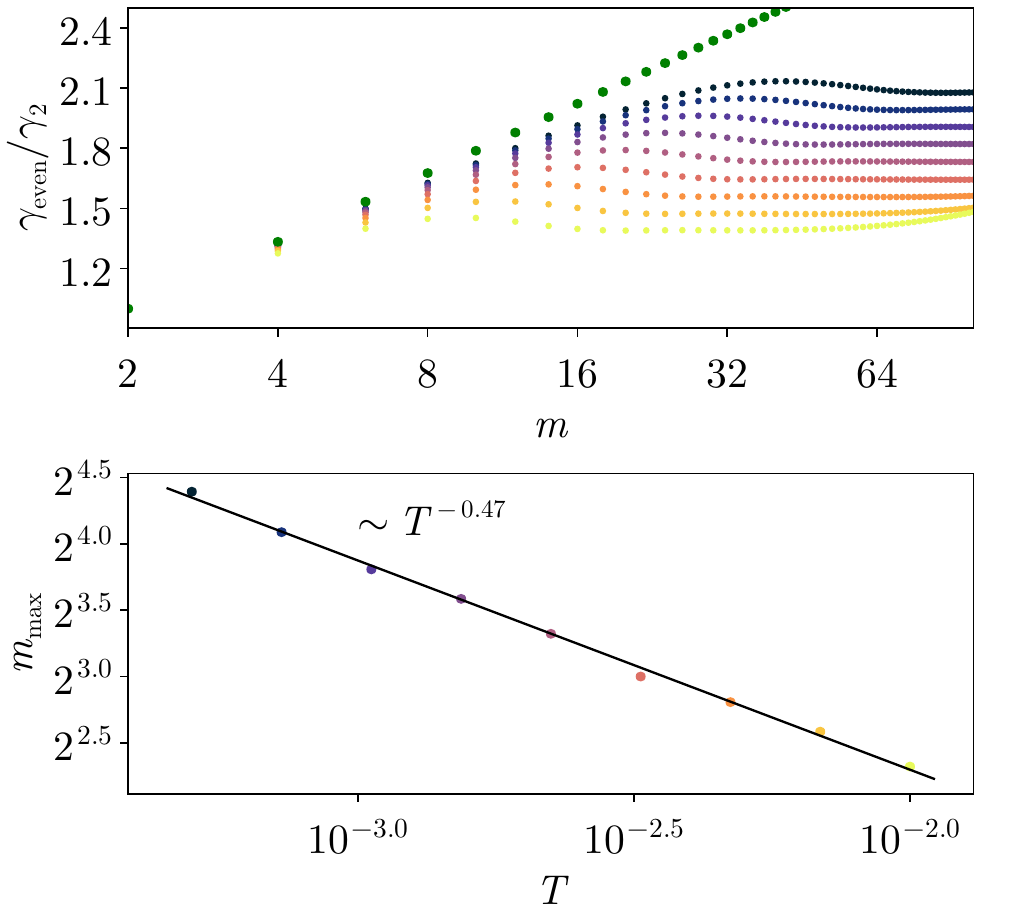}
    \caption{Semi-log plot of the even spectra with the low temperature analytical model shown in green. 
    As temperature is lowered, the lowest even modes collapse onto the model spectrum over a region which grows as $\sim 1 / \sqrt{T/T_F}$.}
    \label{fig:even_spectrum_fit}
\end{figure}

The even rates in \autoref{fig:even_spectrum_fit} shown plotted against the logarithm of $m$ are linear over a region which scales as $\sim 1 / \sqrt{T/T_F}$. 
As temperature is lowered, we find improved agreement to the low-temperate analytical model spectrum from \cite{NilssonEtAl2025}, whose asymptotic form (discussed in \autoref{appendix:asymptotics}) grows logarithmically with $m$.

\section{Low-temperature model spectrum}
From \citet{NilssonEtAl2025}, the even rates for small $m$ are given by 
\bea\label{eq:full-hofmann-even}
\gamma_{m,\text{even}} &= \frac{8\pi T_F}{3 \hbar} \left (\frac{T}{T_F} \right)^2\frac{|V|^2}{\epsilon_T^2} F[m] \equiv \gamma_{2} F[m]
\eea
for a constant interaction matrix element $V$ and $\epsilon_T$ the thermal de Broglie wavelength, with $F[m] = \sum_{j = 1}^{m /2} \frac{1}{2j - 1}$ a slowly increasing function of $m$ with $F[2]=1$ and $F[m \to \infty] \sim \frac12 \log(m)$ (see \autoref{appendix:asymptotics}).
In the small $m$ region, \citet{NilssonEtAl2025} further found analytically that, for $m\geq3$,
\bea \label{eq:full-hofmann-odd}
\gamma_{m,\text{odd}} = \frac{4 \pi^3 T_F }{15 \hbar} \frac{|V|^2}{\epsilon_T^2} m^4 \left (\frac{T}{T_F} \right)^4 \equiv \gamma_{3} \left(\frac{m}{3}\right)^4.
\eea

\section{Asymptotic Limit of $F[m]$} \label{appendix:asymptotics}
The series $F[m]$ which defines the rise of the even rates is:
\bea
F[m]\equiv\sum_{j=1}^{m/2}\frac{1}{2j-1}
=1+\frac13+\cdots+\frac{1}{m-1}
\eea
for even $m$. This can be written in terms of harmonic numbers as
\bea
\sum_{j=1}^{m/2}\frac{1}{2j-1}
=\sum_{k=1}^{m}\frac{1}{k}-\sum_{j=1}^{m/2}\frac{1}{2j}
=H_m-\frac12 H_{m/2},
\eea
where
$
H_n=\sum_{j=1}^n \frac1j
$
is the $n$th harmonic number. Using the asymptotic form
\bea
H_n=\log n+\gamma+O(n^{-1}),
\eea
with $\gamma$ the Euler--Mascheroni constant, we obtain
\bea
F[m] 
\approx
\log m+\gamma-\frac12\left(\log\frac{m}{2}+\gamma\right)
=\frac12\bigl(\log(2m)+\gamma\bigr).
\eea

\section{Krylov gauge transformation}

Starting from the BTE in the basis of real cylindrical harmonics, $\varphi_{m, \text{odd}} =  \cos(m \theta), \varphi_{m, \text{even}} = \sin(m\theta)$.
\begin{equation}\label{eq:recursion}
    \partial_t \varphi_m
    = (-1)^m \frac{i q v_F}{2}\bigl(\varphi_{m+1}-\varphi_{m-1}\bigr)
    - \gamma_m \varphi_m \, .
\end{equation}

Define a parity-dependent phase (``gauge'') transformation
\begin{equation}\label{eq:gauge}
    \tilde\varphi_m \equiv i^{a_m}\varphi_m,
    \qquad
    a_m \equiv \frac{1+(-1)^m}{2}
    =
    \begin{cases}
        1, & m \ \text{even},\\
        0, & m \ \text{odd}.
    \end{cases}
\end{equation}
Equivalently, $\varphi_m = i^{-a_m}\tilde\varphi_m$. 

Substituting $\varphi_{m\pm 1}=i^{-a_{m\pm 1}}\tilde\varphi_{m\pm 1}$ into \eqref{eq:recursion} and multiplying both sides by $i^{a_m}$ yields
\begin{equation}\label{eq:recursion-tilde-intermediate}
    \partial_t \tilde\varphi_m
    = (-1)^m i^{a_m-a_{m+1}} \frac{i q v_F}{2}\,(\tilde\varphi_{m+1} - \tilde\varphi_{m-1})
      - \gamma_m \tilde\varphi_m.
\end{equation}
Now note 
\begin{equation}\label{eq:phase-identity}
    a_m-a_{m+1}
    = \frac12\Bigl[(1+(-1)^m)-(1+(-1)^{m+1})\Bigr]
    = (-1)^m.
\end{equation}
Therefore,
\begin{equation}
    (-1)^m i^{a_m-a_{m\pm 1}}
    = (-1)^m i^{(-1)^m}
    =
    \begin{cases}
        i, & m \ \text{even},\\
        i, & m \ \text{odd},
    \end{cases}
    = i,
\end{equation}
so the hopping prefactor becomes
\begin{equation}
    \Bigl[(-1)^m i^{a_m-a_{m\pm 1}}\Bigr]\frac{i q v_F}{2}
    = i\cdot \frac{i q v_F}{2}
    = -\frac{q v_F}{2}.
\end{equation}
Plugging this into \eqref{eq:recursion-tilde-intermediate}, we obtain 
\begin{equation}\label{eq:recursion-tilde}
    \partial_t \tilde\varphi_m
    = -\frac{q v_F}{2}\bigl(\tilde\varphi_{m+1}-\tilde\varphi_{m-1}\bigr)
    - \gamma_m \tilde\varphi_m \, .
\end{equation}
which is the form used in the main text.

\section{Spectrum of the Liouvillian $\mathcal{L}_q$ and sum rule}

\begin{figure}[h!]
    \centering
    \includegraphics[width=\linewidth]{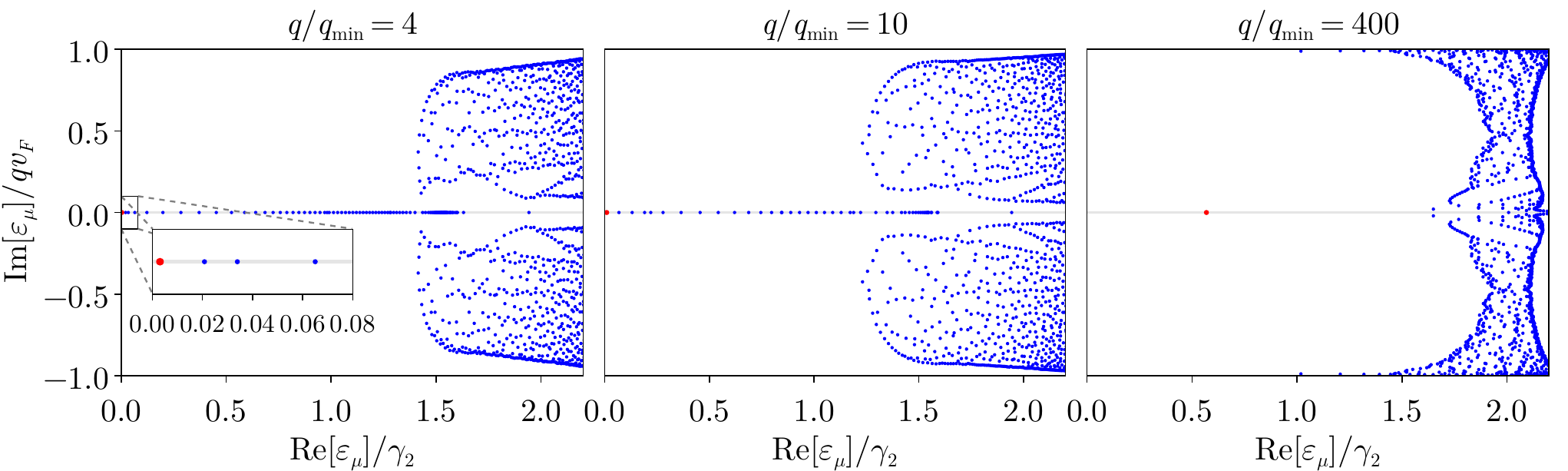}
    \caption{Spectrum of $\mathcal{L}_q$ for $T/T_F = 0.005$ at three values of $q$ in the tomographic regime. 
    The red dot identifies the eigenvalue with minimal real part; i.e., the late-time decay rate of the Krylov wavefunction.
    }
    \label{fig:Lq-spectrum}
\end{figure}

\emph{Spectrum of the Liouvillian.}---
Rewriting Eq.~\eqref{eq:krylov-chain} as $\partial_t |\varphi\rangle = -\mathcal{L}_q |\varphi\rangle$ defines the Liouvillian $\mathcal{L}_q \equiv L + i \q \cdot \v$.
Here we show the spectrum of the Liouvillian (eigenvalues $\varepsilon_\mu$) obtained numerically for various values of $q$.
The mode with the slowest decay rate is shown in red, and is the one giving the low-frequency hydrodynamic pole discussed in the main text.
Note that these spectra are obtained by diagonalizing the full Liouvillian including both even and odd modes.

As explained in the main text, after integrating out analytically the even modes and taking the continuum limit, at ``low-energy'' the Liouvillian maps to a Schr\"odinger operator for a quartic oscillator, $- \psi^{\prime\prime}(x) + \frac{1}{2}|x|^4 \psi(x) = E\psi(x)$ under reflecting boundary conditions at $x = 0$. 
The two lowest eigenvalues of $\mathcal{L}_q$ are therefore predicted to correspond to the two lowest \emph{even parity} eigenvalues of the quartic oscillator on the whole real line. 
These energies have to be solved numerically \cite{quartic-oscillator}, and it can be shown that $E_0 = 0.420805, E_2 = 2.95886$, such that $E_2 / E_0 \approx 7.03$.
This prediction is well verified numerically: At $q = 10 q_\text{min}$, the ratio of the two lowest eigenvalues is 7.04, and this ratio holds approximately across the entire tomographic plateau. This separation ensures that at late times (or low frequencies), the dynamics of the Krylov wavefunction is indeed dominated by a single quasinormal mode for all $q$ in the tomographic regime. 

Overall, Fig.~\ref{fig:Lq-spectrum} confirms the validity of the Schr\"odinger operator picture to describe the Liouvillian spectrum close to the imaginary axis, $\mathrm{Re}[\epsilon_\mu] \ll \gamma_2$.
For larger decay rates, $\mathrm{Re}[\epsilon_\mu] \gtrsim \gamma_2$, one can observe a dense spectrum with $\mathrm{Im}[\epsilon_\mu] \neq 0$, indicative of ballistic modes.

\emph{Sum rule.---}
The real part of the conductivity is subject to the sum rule $\int_0^\infty d\omega\,\mathrm{Re}[\sigma(q,\omega)] = \frac{\pi}{2}$, which can be understood as follows. From the Liouvillian, the conductivity is given by $\sigma(q,\omega) = \langle j_x | (i\omega + \mathcal{L}_q)^{-1} | j_x \rangle$, where $|j_x\rangle$ is the initial current state localized at $m=1$. Evaluating this in the biorthogonal eigenbasis of $\mathcal{L}_q$ (with eigenvalues $\epsilon_\mu$ and right/left eigenvectors $|R_\mu\rangle$, $\langle L_\mu|$) yields a sum of dissipative poles, $\sigma(q,\omega) = \sum_\mu Z_\mu (i\omega + \epsilon_\mu)^{-1}$, with spectral weights $Z_\mu = \langle j_x | R_\mu \rangle \langle L_\mu | j_x \rangle$. Because $\mathrm{Re}(\epsilon_\mu) > 0$, frequency integration gives $\int_0^\infty d\omega\,\mathrm{Re}[\sigma(q,\omega)] = \frac{\pi}{2} \sum_\mu Z_\mu = \frac{\pi}{2} \langle j_x | j_x \rangle = \frac{\pi}{2}$.

\section{ AC current in a channel}
In this Appendix, we give more details on the AC conductance in a channel of width $W$, with no-slip boundary conditions.
We first give the main results in \ref{AppChannelRes}, and then show the derivations in \ref{AppChannelDer}.

\subsection{Results}
\label{AppChannelRes}

\begin{figure}[h!]
    \centering
    \includegraphics[width=0.5\linewidth]{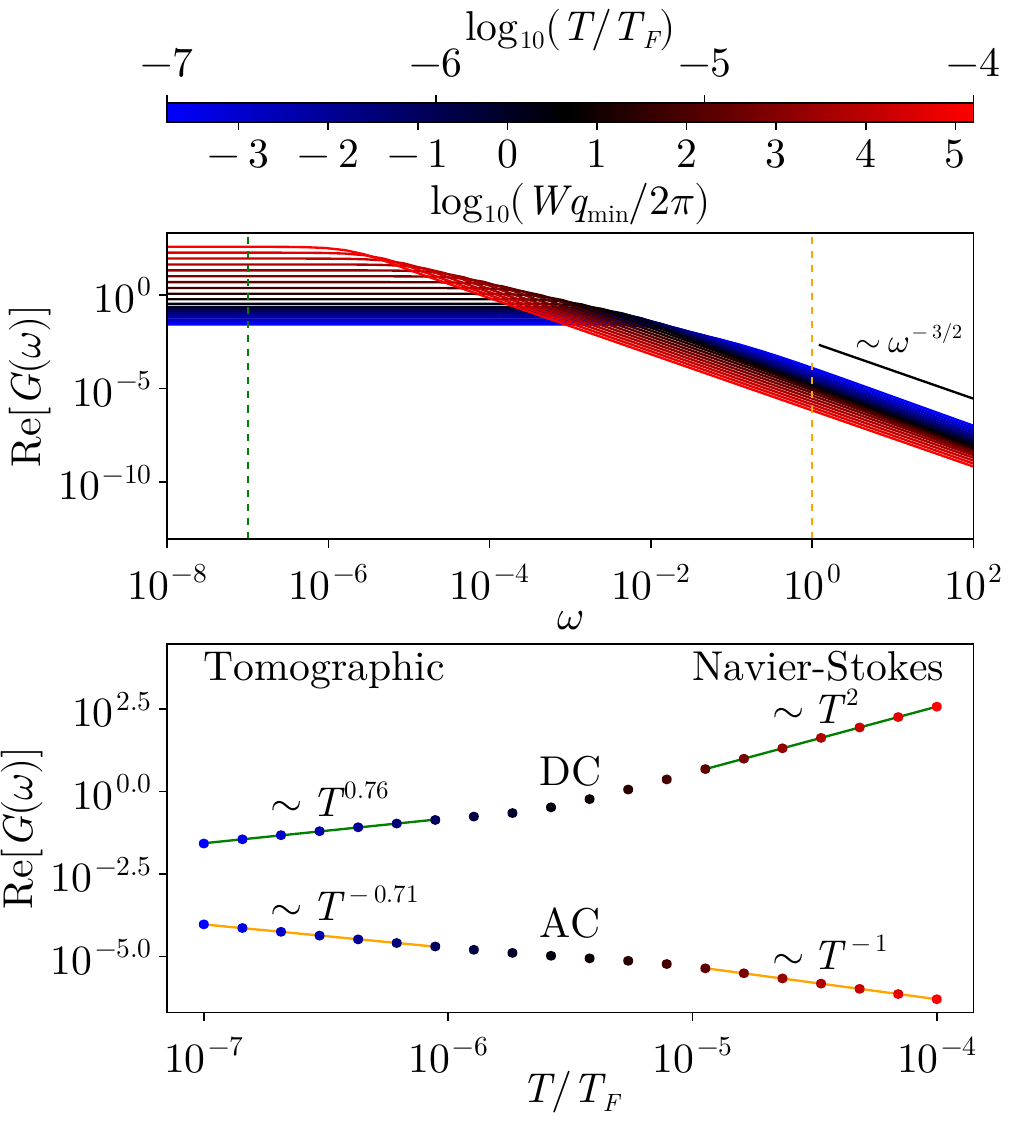}
    \caption{Top: channel conductance as a function of $\omega/\qmin(T_0)$ for fixed width $W=0.001\,\qmin^{-1}(T_0)$, where $T_0/T_F=10^{-7}$. Blue and red respectively denote temperatures with $\qmin(T)<2\pi/W$ and $\qmin(T)>2\pi/W$. Bottom: temperature dependence of the conductance at low (DC) and high (AC) frequencies (the two vertical lines in top panel show the two frequency values). The Navier--Stokes regime, at higher $T/T_F$, shows the expected $T^2$ scaling at low frequency and $T^{-1}$ scaling at high frequency. The tomographic range (smaller $T/T_F$) shows a different scaling with temperature, with extracted scaling $T^{0.76}$ for DC and $T^{-0.71}$ for AC. These exponents are lower than the ones predicted deep in the tomographic regime ($T^{+1}$ scaling for DC and $T^{-1}$ for AC) due to the slow drift of exponents.}
    \label{fig:conductance}
\end{figure}

\emph{Channel-conductance scaling.}---
Here we discuss the dependence of the channel conductance on temperature, width, and frequency; see Fig.~\ref{fig:conductance}. In that figure we fix the width and vary $T/T_F$, allowing us to scan from the Navier--Stokes regime at higher temperatures to the tomographic regime at lower temperatures. In the low-frequency limit, the relevant distinction is whether $W\qmin\gg 1$ (Navier--Stokes) or $W\qmin\ll 1$ (tomographic). In the first case, one finds
\bea
\frac{G}{W}\sim \frac{W^2}{\eta}\propto T^2
\qquad\text{(Navier--Stokes)},
\eea
whereas in the tomographic regime
\bea
\frac{G}{W}\sim \frac{\qmin^\alpha W^{z+\alpha}}{\eta_\star}
\propto T^{3(z+\alpha)-4}.
\eea
Deep in the asymptotic tomographic regime, $z+\alpha\to 5/3$, recovering the linear-in-$T$ conductance predicted in Ref.~\cite{KryhinEtAl2025}. In practice, however, the logarithmic drift of the exponents remains substantial even at ambitious values of $T/T_F$ and $W\qmin$, as seen in the lower panel of Fig.~\ref{fig:conductance}, where the numerical DC conductance scales approximately as $T^{0.76}$. This may be important when comparing quantitatively with experiments extracting viscosities, such as Ref.~\cite{ZengEtAl2024}.

At higher frequency, the conductance behaves parametrically as
\bea
\mathrm{Re}\,G &\sim \frac{\delta}{\omega}(\qmin\delta)^{-\alpha}, \\
\mathrm{Im}\,G &\sim \frac{W}{\omega}.
\eea
These expressions work in both regimes: 
in the Navier--Stokes regime ($ \delta \gg \qmin^{-1}$), where $\alpha=0$ and $\delta=\sqrt{\eta/\omega}$, and the tomographic regime ($ \delta \ll \qmin^{-1}$), where $\alpha=1/3$ and $\delta=(\eta_\star/\omega)^{3/4}$. As expected, $\mathrm{Re}\,G$ becomes independent of $W$ at high frequency because only the boundary layers contribute appreciably to the in-phase current.

In the Navier--Stokes regime this gives
\bea
\mathrm{Re}\,G(\omega)\sim \frac{1}{\omega^{3/2}\sqrt{\gamma_2}}\propto T^{-1},
\eea
whereas in the tomographic regime we find
\bea
\mathrm{Re}\,G(\omega)
\sim
\frac{1}{
\omega^{\,\frac{p_\mathrm{eff}}{2p}+1}
\sqrt{\gamma_2}\,
\gamma_3^{-\frac{p_\mathrm{eff}}{2p}+\frac12}
}
\propto
T^{-3+2p_\mathrm{eff}/p},
\eea
with $p_\mathrm{eff}/p=2(1-\alpha(q))/z(q)$ encoding the drift of the apparent exponents. Asymptotically, $p_\mathrm{eff}\to p$ and the high-frequency temperature dependence approaches the same $T^{-1}$ scaling as in the Navier--Stokes regime, even though the spatial structure of the current remains very different. The frequency scaling is confirmed in the top panel of Fig.~\ref{fig:conductance}; the lower panel shows that the temperature scaling in the tomographic regime still exhibits substantial drift, with an observed behavior close to $T^{-0.71}$ for the parameters used there.

\subsection{Derivations}
\label{AppChannelDer}
\emph{Current profiles}.---
We consider transport in an infinite 2D channel, $0 < x < W$, where the current is driven by an electric field $\mathbf{E} = E e^{i\omega t}\hat{y}$. 
From \cite{LedwithEtAl2019}, extended to finite frequency, the current profile for no-slip boundary conditions reads
\bea\label{eq:current-profile}
    j(x) &= E \frac{\sum_{n \neq 0} \sigma(q_n, \omega) - \sum_{n \neq 0} \sigma(q_n, \omega) e^{iq_n x}}{1 + \sum_{n \neq 0} \sigma(q_n, \omega) / \sigma(0, \omega)} 
\eea
with $q_n = 2\pi n / W$, and $\sigma(0,\omega) = 1/(i \omega + \gamma_\text{mr})$.
We used this formula to calculate numerically the current profiles shown in Fig 5 of the main text.

\emph{Channel conductance scaling}.---
The channel conductance, $G(\omega)$, can be obtained by integrating over $x$ giving:
\bea\label{eq:conductance}
    G(W, \omega) = \frac{W}{i \omega + \gamma_\text{mr} + (\sum_{n \neq 0} \sigma(q_n, \omega))^{-1}}
\eea
where $\gamma_\text{mr}$ denotes the scattering rate corresponding to momentum-relaxing collisions which we will take to be zero in the rest of this appendix.

One can now ask what the asymptotic scaling of $G(\omega)$ is for the single pole approximation of the conductivity, $\sigma(q, \omega) = (q/\qmin)^{-\alpha} /(i \omega + \eta_\star q^z)$. Let $S = \sum_{n > 0} \sigma(q_n, \omega)$ and define the dimensionless frequency $\Omega = \omega (W/2\pi)^z / \eta_\star$. We find
\bea
    S &= \left(\frac{W}{2\pi}\right)^{\alpha + z} \frac{\qmin^\alpha}{\eta_\star}  \sum_{n > 0} \frac{n^{-\alpha}}{i\Omega + n^z}\\
\eea
In the DC limit $\Omega \to 0$, the sum is just a numerical constant, and thus
\bea
    \text{Re}[G(W, \omega = 0)] &= W \text{Re}[S]\\
    &\sim W^{1 + \alpha + z} \qmin^{\alpha} \eta_\star^{-1}
\eea
In the Navier-Stokes regime ($\alpha=0, z=2, \eta_\star=\eta\sim\gamma_2^{-1}$), this specifies to
\bea
\text{Re}[G(W, \omega = 0)] &\sim W^{3} \eta^{-1} \propto W^3 T^2
\eea
For the tomographic regime, the $T$ scaling then reads
\bea
    \text{Re}[G(W, \omega = 0)]  &\propto W^{1 + \alpha + z} T^{3(z + \alpha) - 4}
    \eea
where we used the fact that $\eta_\star \sim T^{4 - 3 z}$.

In the high frequency limit, the summand becomes negligible above $n_\text{max} \sim |\Omega|^{1/z}$. Thus the magnitude of $S$ scales as:
\bea
    |S| &\sim \left(\frac{W}{2\pi}\right)^{\alpha + z} \frac{\qmin^\alpha}{\eta_\star} n_\text{max}\left[\frac{n_\text{max}^{-\alpha}}{n_\text{max}^z}\right]\\
    &\sim \left(\frac{W}{2\pi}\right)^{\alpha + z} \frac{\qmin^\alpha}{\eta_\star} |\Omega|^{(1 - \alpha - z) / z}\\
   &\sim \frac{\qmin^{\alpha}}{\eta_\star^{(1 - \alpha)/z}} \left(\frac{W}{2\pi}\right) \omega^{\frac{1 - \alpha}{z} - 1}
\eea
Note that $S$ has both a real and imaginary parts that of the same order.

In the high frequency limit, the scaling of the real part of the conductance is then given by
\bea
    \text{Re}[G(\omega)] &\approx W\frac{\text{Re}[S^{-1}]}{\omega^2} \qquad \text{for $\Omega \gg 1$}\\
    &\sim \qmin^{- \alpha}\eta_\star^{\frac{1 - \alpha}{z}} \omega^{- \frac{1 - \alpha}{z} - 1} 
\eea

In the Navier-Stokes regime, this simplifies to
\bea
    \text{Re}[G(\omega)] \sim \eta^{\frac1{2}} \omega^{- 3/2} \propto \frac1{T \omega^{3/2}}
\eea

In the tomographic regime, one finds instead
\bea
\text{Re}[G(\omega)] \sim \qmin^{-1} \gamma_3^{\frac{1 - \alpha}{z}} \omega^{- \frac{1 - \alpha}{z} - 1} 
    \propto T^{-3 + 4\frac{1 - \alpha}{z}} \omega^{- \frac{1 - \alpha}{z} - 1}
    \eea

The scaling with temperature is summarized in \autoref{tab:T-scaling}.

\begin{table}[t!]
\centering
\begin{tabular}{c|cc}
  & Viscous & Tomographic \\ \hline
$\omega = 0$ & 2 & $3 (z +\alpha) - 4$ \\
$\omega \to \infty$ & -1 & $-3 + 4(1 - \alpha)/z$ \\
\end{tabular}
\caption{\label{tab:T-scaling}Temperature scaling exponents of $\text{Re}[G(\omega)]$.}
\end{table}

\end{document}